\documentclass[11pt,a4paper]{article}
    \usepackage{jheppub}
\usepackage{amsfonts,amssymb,amsthm,array,amsmath,mathrsfs}
\usepackage{relsize}
\usepackage{cancel}
\usepackage{mathtools}
\usepackage[T1]{fontenc}
\usepackage{fancyhdr}
\usepackage{hyperref}
\usepackage{bibentry}
\usepackage[colorinlistoftodos]{todonotes} 

\usepackage{graphicx}
\usepackage{dcolumn}
\usepackage{bm}
\usepackage{enumerate}
\usepackage{slashed}
\usepackage{simplewick}
\usepackage{comment}

\usepackage{soul}

\usepackage[english]{babel}
\usepackage{appendix}
\usepackage[utf8]{inputenc}

\usepackage{marginnote}
\usepackage[normalem]{ulem}

\usepackage{subfigure}
\newcommand{\sA}{\scriptscriptstyle\rm A}

\newcommand{\nn}{\nonumber}

\newcommand{\eq}[1]{(\ref{#1})}

\newcommand{\la}{\label}
\newcommand{\ba}{\begin{align}}
\newcommand{\ee}{\end{equation}}
\newcommand{\be}{\begin{equation}}

\def\12{\frac{1}{2}}
\newcommand{\p}{\partial}
\newcommand{\en}{\end{align}}
\newcommand{\e}{\epsilon}

\usepackage{color}

\usepackage{amsmath}

\title{Chiral Anomaly in Euler Fluid and Beltrami Flow }

\author[a]{P.B.~Wiegmann}
\author[b]{and A.G.~Abanov}

\affiliation[a]{Kadanoff Center for Theoretical Physics, University of Chicago,
5640 South Ellis Ave, Chicago, IL 60637, USA}

\affiliation[b]{Department of Physics and Astronomy and Simons Center for
Geometry
and Physics, Stony Brook University, Stony Brook, NY 11794, USA}

\emailAdd{wiegmann@uchicago.edu}
\emailAdd{alexandre.abanov@stonybrook.edu}

\abstract{We show that the chiral anomaly of quantum field theories with Dirac fermions subject to an axial background field is an inherent property of kinematics of a perfect classical fluid. Celebrated Beltrami flows (stationary solutions of Euler equations with extensive helicity) exhibit the chiral anomaly equivalent to that known for Dirac fermions. A prominent effect of the chiral anomaly is the transport electric current at equilibrium. We show that it is also a property of Beltrami flows.}

\begin{document} 

\maketitle\flushbottom

\section{Introduction} 

Interacting quantum fermionic systems can behave as perfect fluids. Prominent examples are superconductors, the superfluid \(^3\hbox{He}\), the fractional quantum Hall effect, and one-dimensional electronic systems, e.g., Luttinger liquid. Recently a possibility that a quark-gluon plasma arising in collisions of heavy ions may turn into a fluid came to light. The fundamental particles in quark-gluon plasma, \(^3\hbox{He}\)  and some other examples are Dirac fermions. A characteristic feature of  quantum field theories (QFT) with Dirac fermions is quantum anomalies that are mainly insensitive to interactions. Therefore, if the Dirac fermions turn into a fluid, anomalies must be an inherent part of hydrodynamics as well. This argument set off a search for relativistic hydrodynamics models consistent with the anomalies (see, Refs.~\cite{son2009hydrodynamics,haehl2014effective}, and references therein). In a recent paper, \cite{abanov2022axial},  the effect of the axial-current anomaly had been identified in the foundational description of fluid dynamics  -- the conventional  \underline{classical Euler equations} (relativistic and non-relativistic alike).  
    
If the Dirac fermions retain chiral imbalance, a fermionic system exhibits a physically significant anomaly-driven effect: an electric current runs across the system at the thermodynamic equilibrium. This work shows that this and other anomaly-based properties are readily apparent in Beltrami flows of the perfect classical fluid. We recall that the Beltrami flows are stationary flows with extensive helicity (i.e., proportional to the volume of the system). Superfluid \(^3\hbox{He-A}\) may serve as a platform for the experimental realization of the effects discussed in this work. There is an experimental evidence of both the axial current anomaly \cite{bevan1997momentum}, and of realization of the Beltrami flow \cite{salomaa1987quantized}.

\subsubsection*{Chiral anomaly in Dirac fermions} We briefly review   the chiral anomaly in the fermionic quantum field theories before turning to hydrodynamics.
Dirac fermions, whose Hamiltonian density is
\begin{align}
    H=\psi_{\scriptscriptstyle L}^\dag\bm\sigma(i\bm\nabla+\bm A)\psi_{\scriptscriptstyle L} -\psi_{\scriptscriptstyle R}^\dag\bm\sigma(i\bm\nabla +\bm A) \psi_{\scriptscriptstyle R}
 \la{0}
\end{align}
in a time-independent magnetic field \(\bm B=\bm\nabla\!\times\! \bm A\)
possess two conserved charges. They are the electric charge \(Q\) and the axial charge referred to as the chirality \( Q_{\sA}\)  
\be 
    Q=\int \rho d\bm x\bm=N_{\scriptscriptstyle L}+N_{\scriptscriptstyle R},\,\quad Q_{\sA}
    =\int \rho_{\sA} d\bm x=N_{\scriptscriptstyle L}-N_{\scriptscriptstyle R} \,.
 \la{1}
\ee 
The densities are the sum \(\rho=\psi^\dag_{\scriptscriptstyle L} \psi_{\scriptscriptstyle L}+\psi^\dag_{\scriptscriptstyle R}\psi_{\scriptscriptstyle R}\) and the difference \(\rho_{\sA}=\psi^\dag_{\scriptscriptstyle L} \psi_{\scriptscriptstyle L}-\psi^\dag_{\scriptscriptstyle R}\psi_{\scriptscriptstyle R}\) of the left and the right occupation numbers of chiral (Weyl) components of the Dirac multiplet \((\psi_{\scriptscriptstyle L},\, \psi_{\scriptscriptstyle R})\).
They are  temporal components of the 4-vector  current  \(j=(j^0,\bm j)\) and 4-axial current \(j_{\sA}=(j_{\sA}^0,{\bm  j}_{\sA})\). 

As a system with two conserved charges, Dirac fermions could be brought in contact with a reservoir which maintains the chemical potential \(\mu\) coupled to the particle density \(\rho\) and the {\it chiral chemical potential} \(\mu^{\sA}\) coupled to the chirality density \(\rho_{\sA}\). Such a system may reside in thermodynamic equilibrium with the reservoir. Then microstates are described by a grand canonical ensemble with the Gibbs probability \(e^{-\mathscr{H}/T}\) with 
\begin{align}
    \mathscr{H} = H-\!\int(\mu\rho+\mu^{\sA}\rho_{\sA})\, d\bm x\,,
 \la{4}
\end{align}   
where \(H\) is the energy of the isolated system. The chemical potential \(\mu\) may be treated as a temporal component of the external electromagnetic 4-vector potential \(A=(\mu,\bm A)\) and the chiral chemical potential \(\mu^{\sA}\) as  a temporal component of the external axial 4-vector potential  \(A^{\sA}=(\mu^{\sA},\bm A^{\sA})\) commonly considered in the Dirac QFT. 
In the main part of the paper we focus on the case \(\bm A^{\sA}=0\). The general case is briefly considered in Sec.~\ref{flows}. We consider a case of zero temperature, where the equilibrium state is the ground state of \(\mathscr{H}\). There  \(\mu\) and \(\mu^{\sA}\) determine the charges \(Q\) and \(Q_{\sA}\). 

In the familiar case, when the chemical potential is coupled to the particle number, the equilibrium state assumes no flow. The situation is fundamentally different when the reservoir maintains the chiral chemical potential \(\mu^{\sA}\). In this case, the equilibrium  state carries an electric transport current proportional to and directed along the magnetic field 
\begin{align}
    \bm j_{\rm trans}=k\mu^{\sA} \bm B\,.
 \la{61}
\end{align}    
This result had a tangled history and met interpretational difficulties.
Formally one could extract it from early works on anomalies \cite{bardeen1969anomalous,bardeen1984consistent} where fermions are considered in the background of the axial vector potential. Physics applications started perhaps from the work of Vilenkin \cite{vilenkin1980equilibrium}. Recently \eq{61} had been explored in heavy-ion collisions \cite{kharzeev2008effects} among other fields. A clear interpretation of Eq.~\eq{61} as a ballistic transport at equilibrium is presented in Ref.~\cite{alekseev1998universality} (see, also \cite{frohlich2000new}. In one spatial dimension  the analog of Eq.~\eq{61} is the Landauer formula   \(\bm j_{\rm trans}= k\mu^{\sA}\)   \cite{landauer1970electrical,buttiker1986four} for  the experimentally observed Landauer-Sharvin effect of the resistance of ballistic transport. There \(\mu^{\sA}\) is a voltage drop across the bulk and the constant \(k\) is the conductance quantized in fundamental units \(\text{}e^2/h\).  The effect of the transport current at equilibrium also discussed in connection with  non-centrosymmetric semiconductors and cosmology. For some references and applications of the effects of the chiral reservoir in various domains of physics, see Refs.~\cite{chernodub2021thermal,rogachevskiilaminar,Avdoshkin2016Onconsistency}.

The origin of this result is the axial-current anomaly obtained by Adler \cite{adler1969axial} and Bell and Jackiw \cite{bell1969pcac} in 1969 (see also Ref.~\cite{treiman2014current}). That is, under a physics requirement that the vector current is conserved in an isolated system, the divergence of the axial current is necessarily given by the formula
\begin{align}
    \p\!\cdot\! j_{\sA}=k\bm E\!\cdot\! \bm B,\quad k=2 \,.
 \la{31}
\end{align}
The coefficient \(k\) is the value of the triangle diagram with one axial and two vector vertices. In units of the magnetic flux quantum \(\Phi_0=hc/e\),
the coefficient \(k\) is equal \(2 \), the number of Weyl components in the Dirac multiplet. Then, if the system is coupled to reservoirs, the  same triangular diagram which entered \eq{31}, determines that a gradient of the chiral chemical potential causes the vector current to diverge 
\begin{align}{
  \p\!\cdot\! j=k\bm E^{\sA}\!\cdot\!\bm B\,,
  \quad  \bm E^{\sA}:=\bm\nabla\mu^{\sA}\,.
 \la{51}}
\end{align}
Eq.~\eq{61} is a particular stationary solution of the chiral anomaly equation \eq{51}. We comment that Eq.~\eq{61} is valid for \(\mu^{\sA}={\rm const}\) even though the r.h.s. of \eq{51} vanishes.\footnote{Throughout the paper we use a dot to denote the contraction of adjacent indices, i.e., \(\p\cdot\! j\) denotes \(\p_\alpha j^\alpha =\dot\rho+\bm\nabla \!\cdot\bm j\).} 

\subsection*{Helicity and Beltrami flow of a perfect fluid}

Now we turn to the electrically charged barotropic perfect (Euler's)  fluid governed by the Hamiltonian 
\begin{align}
 \la{6}
    H=\int\left(\tfrac 12m\rho\bm v^2+\varepsilon[\rho]\right) d\bm x\,.
\end{align}
Here \(\rho,\bm v  \) are the fluid density and  velocity,  \(m\) is the
mass of fluid particles and  the energy density \(\varepsilon\) is a function of \(\rho\).

Like  Dirac fermions,  Euler flows in three dimensions possess two conserved charges. One is the particle number \(Q=\int \rho d\bm x\). The second conserved charge is the {\it fluid helicity}
\begin{align}
\mathcal{H}
    =h^{-2}\int m\bm v\!\cdot\! \bm \omega\,d\bm x\,.
 \la{7}
\end{align} 
Here, \(\bm\omega=\bm\nabla\!\times\! m\bm v\) is the fluid vorticity, and \(h\) is a normalization constant of the dimension of \(\text{\it energy}\times\text{\it time}\) which makes helicity dimensionless. The root of the helicity conservation can be traced to its topological nature: helicity as a linkage of vortex lines \cite{moffatt1969degree}. Another related root is the degeneracy of the Poisson structure of the Euler hydrodynamics. The latter gives helicity the meaning of the Casimir invariant (see, e.g., \cite{zakharov1997hamiltonian}). It is conserved regardless of the choice of the Hamiltonian. 

For the reasons described above, the fluid could  be coupled with two reservoirs as in \eq{4}. One maintains the particle density (the electric charge), the second maintains the helicity 
\begin{align}\la{190}
   \mathscr{H}= \int\left[\tfrac 12m\rho\bm v^2+\varepsilon[\rho]-\mu\rho -\mu^{\sA}m\bm v\!\cdot\!\bm\omega\right] d\bm x\,.
\end{align}
In the next section we  explain how to add the electromagnetic field to the chiral coupling. Then we will show that the formulas for anomalies (\ref{31},\ref{51},\ref{61}) hold to the Euler fluid. Furthermore, if the normalization constant \(h\) in \eq{7} is identified with the Planck constant, the formulas become identical. With this in mind, we will use the units of \(h\) and \(\Phi_0\), setting $h=\Phi_0=1$ in \eq{190} and below.

We focus on  the electric transport current present in the equilibrium state \eq{61}. From the fluid perspectives the equilibrium state is a stationary (or steady) flow.  We will show that the chiral reservoir yields a stationary flow broadly known as a {\it Beltrami flow} \cite{arnold2008topological}. We recall that the (generalized) Beltrami flow is defined by the condition that velocity and vorticity are collinear at every point
\be\la{Beltrami}
    {\rm curl}\,\bm v=\kappa\bm v\,.
\ee 
Beltrami flows carry an extensive helicity, which, as we will show, itself is determined by $\mu^{\sA}$. Then we show that the Beltrami flow conducts an electric current given by the formulas identical to the formulas (\ref{61},\ref{51}) of the fermionic QFT.

\subsection*{Axial current anomaly in fluid dynamics and chiral coupling}

In magnetic field the definition of helicity amounts to replacing the momentum \(m\bm v\) in \eq{7} by the canonical momentum 
\be 
    m\bm v\to\bm\pi:=m\bm v+\bm A\,, 
 \la{8}
\ee
where \(\bm A\) is the  vector potential so that \(\bm B=\bm\nabla\!\times\! \bm A\). Under this definition the helicity 
\be
    \mathcal{H}=\int \bm\pi\cdot\bm\nabla\!\times\! \bm\pi \,d\bm x \la{90}
\ee 
remains conserved \cite{bekenstein1987helicity,khesin1989invariants} (see also the section VI.2.C of \cite{arnold2008topological}). However, the helicity  \(\mathcal{H}\) is no longer  an additive quantity, as its density \(\bm\pi\cdot\bm\nabla\!\times\! \bm\pi\) is not  a local Eulerian field. This creates the difficulty of coupling helicity to the reservoir.  
The problem had been addressed in \cite{abanov2022axial}. There we argued  that in magnetic field the analog of the axial charge is the fluid chirality \(Q_{\sA}=\int\rho_{\sA} d\bm x \), not the helicity. The chirality density $\rho_{\sA}$ is a local Eulerian field
\begin{align}
   \rho_{\sA} =m\bm v\!\cdot\!\left(\bm\omega +2\bm B\right)\,,
 \la{9}
\end{align}
and, therefore is an additive quantity which at \(\bm B=0\) is equal to the helicity density.\footnote{Heuristically, one could arrive at \eq{9} in the following way. In a local frame rotating with a fluid parcel, the magnetic field modifies the angular velocity \(\bm\omega/(2m)\) of a charged parcel by the Larmor frequency \(\bm B/m\). This effectively changes vorticity as \(\bm\omega\to\bm\omega+2\bm B\).} 
It is not difficult to find the relation between chirality and helicity. Their difference depends only on external fields and is equal to the twice magnetic helicity
\begin{align}
    Q_{\sA}=\mathcal{H}-2\int\bm A\!\cdot\!\bm B \,d\bm x\,.
 \la{10}
\end{align}
Based on this arguments we conclude that the chirality density $\rho_{\sA}$ given by \eq{9}, not the helicity density \eq{90} which enters {to the augmented energy} \eq{4}. However, there is a caveat. There are  certain space-time configurations of the electromagnetic field which may change the magnetic helicity, and therefore change the  fluid chirality \eq{10}. It follows from \eq{10} that  
\be\la{Q}
	\tfrac{d}{dt}Q_{\sA}=2\int\bm E\!\cdot\!\bm B \,d\bm x\,.
\ee
The local form of this equation 
\begin{align}
   \p_t{ \rho_{\sA}}+\bm\nabla\!\cdot  \bm j_{\sA}=2{\bm E}\!\cdot\!\bm B\,
 \la{144}
\end{align}     
is the axial-current anomaly \eq{31} [see \eq{2500} below for the explicit formula for the axial flux \(\bm{j_{\sA}}\)]. Like in the case of fermions the chiral reservoir is coupled to a charge whose conservation is subject to the anomaly. This is the source of the effects we study.

Perhaps the most interesting and relevant case for applications occurs at \(\mu,\mu^{\sA}={\rm const}\). In this case \(\mathscr{H}=H-\mu Q-\mu^{\sA} Q_{\sA}\) and one may think about the reservoir as contacts at the system boundaries. Still, we assume that the chemical potentials may vary in space as that clarifies computational and structural aspects of hydrodynamics and the anomalies. Furthermore,  anomaly-based properties  are insensitive to the space-time structure, Galilean or Minkowski. We choose to describe a non-relativistic fluid. In the  end of the text we list the extension of our results  for a general axial potential $A^{\sA}=(\mu^{\sA},\, A^{\sA})$ and present the formulas in a covariant form readily applicable to relativistic fluid dynamics. Similarities between anomalies in QFT and the fluid dynamics also hold in neutral rotating systems. In this case, the angular velocity is identified with the Larmor frequency \(\bm B/m\) and the electric field with the gradient of an electrochemical potential or external forces.

From the perspective of fluid dynamics, fluid motion subject to the chiral reservoir is an interesting problem on its own. To the best of our knowledge, the hydrodynamics in such a setting has not been addressed so far. We develop it below within the Hamiltonian description of the fluid.

\section{The Hamiltonian and the Poisson algebra}

First, we express the fluid chirality density \eq{9} in terms of the canonical momentum $\bm\pi$ \eq{8}   
\begin{align}
    \rho_{\sA}=&\left(\bm \pi-\bm A)\!\cdot\!(\bm\nabla\times\bm\pi+\bm B\right)\,
 \la{131}
\end{align}
and use it in the augmented Hamiltonian \eq{4} 
\begin{align}
    \mathscr{H}[\rho,\bm\pi]=&\int \left[\tfrac \rho{2 m}(\bm\pi\!-\!\bm A)^2 \!+\e[\rho]-\mu\rho-\mu^{\sA}\rho_{\sA}\right] d\bm x\,.
 \la{13}
\end{align}  
The canonical momentum and the fluid density generate the Lie-Poisson  algebra
\begin{align}
    &\{\rho(\bm x),\,\rho(\bm x')\}=
    0\,,\nn\\
   & \{\rho(\bm x),\,\bm\pi(\bm x')\}=
    \bm\nabla_{\bm x'}\delta(\bm x\!-\!\bm x')\,,\la{160}\\
    &\{\bm \pi(\bm x)\times\bm\pi(\bm x')\} =
 (2/\rho)(\bm\nabla\!\times\!\bm\pi)\,\delta(\bm x\!-\!\bm x') \,.\nn
\end{align}
Throughout the paper we routinely drop the argument $\bm{x}$ when it does not compromise the meaning. The brackets (\ref{160}) yield the bracket for the chirality density
\begin{align}
    &\{\rho_{\sA}(\bm x), \rho(\bm x')\}=(\bm\nabla\!\times\!\bm\pi+\bm B)\cdot\bm\nabla_{\bm x'}\delta(\bm x\!-\!\bm x')\,.
 \la{333}
\end{align}
 For completeness we write down the remaining brackets also following from \eq{160} 
\begin{align}
    &\{\bm\pi(\bm x), \rho_{\sA}(\bm x')\}\!
    =-\rho^{-1}\!(\bm\nabla\!\times\!\bm\pi)\times\!(m\bm v\!\times\!\bm\nabla_{\bm x'})\delta(\bm x\!-\!\bm x')\,,
 \la{252} \\
    &\{\rho_{\sA}(\bm x), \rho_{\sA}(\bm x')\}
    =-\Big[\bm\nabla\!\times\!\Big( {\rho}^{-1}(\bm\nabla\!\times\!\bm\pi)\!\cdot\! m\bm v\Big) m\bm v\Big]\!\cdot\!\bm\nabla_{\bm x'}\delta(\bm x\!-\!\bm x')\,.
  \la{2520}
\end{align}
The form of the brackets  (\ref{333},\ref{252},\ref{2520}) reflects that the fluid chirality $Q_{\sA}=\int\rho_{\sA}d\bm x$ is the Casimir function of the Poisson algebra, i.e., Poisson-commute with all hydrodynamic fields. 

\subsubsection*{Magnetic field as a central extension}

We notice that the magnetic field prevents the r.h.s. of \eq{333} from being expressed solely in terms of the canonical momentum \(\bm\pi\). 
The magnetic field acts as a {\it central extension} of the Poisson algebra. In Sec.\ref{sec25} we show that the central extension in the Poisson structure of hydrodynamics is essentially equivalent to the formulation of the effect of the anomaly in terms of the divergence of the currents (\ref{31},\ref{51}). Here we comment on how the axial anomaly in the form of \eq{Q} follows from the bracket \eq{333}.  First we write the Hamiltonian equation $\dot Q_{\sA}=\{Q_{\sA},\mathscr{H}\}$ and notice that the only contribution comes from the term $-\int\mu\rho d\bm x$ in \eq{13}. Other terms are divergence and are eliminated by the integration. Then with the help of \eq{333} we compute  
\begin{align}\la{Q1}
  \dot Q_{\sA}=-\{Q_{\sA},\,\int\mu\rho d\bm x\} 
  =2\int  \bm\nabla\mu\!\cdot\!\bm B\, d\bm x\,.
\end{align}
This yields \eq{Q} after identifying   $\bm\nabla\mu$ with the electric field $\bm E$. 

We remark that if  magnetic field possess  monopoles with a charge  $q_{\rm mon}=-\int \bm\nabla\!\cdot\!\bm B \,d\bm x$, then \eq{Q1}  yields the central extension of the Poisson algebra of the charges
\begin{align}
  \{Q,\,Q_{\sA}\}=2q_{\rm mon}\,.  
\end{align}
The same formula holds for Dirac fermions (see, e.g., Refs.~\cite{Grossman1983Dyon,treiman2014current}). It follows from the current algebra of Faddeev   \cite{faddeev1984operator} (see, also \cite{alekseev1998universality}), where the magnetic field also yields the central extension of the algebra of fermionic currents.

\subsection*{Conjugate fields}

We will use the fields conjugate to the density, the momentum, and to
the external electromagnetic vector potential. The first one is defined as     
\begin{align}
    \pi_0:=-\delta \mathscr{H}/\delta\rho|_{\bm\pi}=\Phi+\mu\,,
\la{23} 
\end{align}
where we denoted the Bernoulli function $\Phi$ and the enthalpy $w$ as
\begin{align}
    \Phi=-\tfrac 12m\bm { v}^2- w[\rho]\,,\quad w=d\varepsilon/d\rho\,.
 \label{bernoulli}
\end{align}
The field conjugate to the momentum is the {\it flow vector field}
 \begin{align}
    \bm{\mathcal{J}}={\delta  \mathscr{H}}/{\delta \bm \pi}|_\rho\,,
 \la{161}
\end{align}
whose meaning we will discuss shortly.

Our primary interest is the electric (or vector) current which we define
as a linear response to the electromagnetic vector potential \(\bm A\) (at a fixed fluid momentum \(\bm \pi\)) evaluated at zero electromagnetic field\footnote{We comment that the electric  current in \cite{alekseev1998universality}
is defined as \(\bm{\mathcal{J}}\).} 
\begin{align}
    \bm j &=-(\delta \mathscr{H}/\delta\bm A|_{\bm\pi})|_{\bm A=0}\,.
 \la{19}
\end{align}
We further clarify the definition of the electric current  in Sec.~\ref{LF}
where we discuss the Lorentz force. 

We remark that in an isolated  Galilean invariant fluid, the fluid momentum per particle, the electric current and the vector flow field   \(\bm v\) are  proportional to the fluid's velocity $\bm v$. The chiral reservoir makes these three vector fields (\ref{8},\ref{19},\ref{161}) different.

\section{Flow vector field and the continuity equation} 

Frozen-in fields are fluid substances whose Lie derivatives along a {\it flow vector field} vanish. We denote the flow vector field as  \(\bm{\mathcal{U}}\). The physical meaning of \(\bm{\mathcal{U}}\) is the  true fluid velocity, which is not necessarily equals to \(\bm v\) serving as a notation for the fluid momentum \(\bm p=m\bm v\). Similarly, the  advection flux \(\bm{\mathcal{J}} =\rho\,\bm{\mathcal{U}}\), defined by  Eq.\eq{161} is the true mass flux. We demonstrate it obtaining the continuity equation as the Hamiltonian equation
\be
	\dot \rho(\bm x)=\{ \rho(\bm x),\,\mathscr{H}\}   =\int\{\rho(\bm x),\bm\pi\}\frac{\delta  \mathscr{H}}{\delta \bm \pi} \,,  
\ee
where the integration goes over the omitted argument. Then with the help of Poisson algebra and \eq{161} we obtain that the 4-advection flux   \(\mathcal{J}=(\rho,\bm{\mathcal{J}})\) is divergence-free 
\begin{align}
 \la{20}
    \p\cdot\! \mathcal{J}\equiv\dot\rho+\bm\nabla\!\cdot \bm{\mathcal{\!J}}=0\,.
\end{align}
We see that the flow along the vector field \(\bm{\mathcal{U}}\) advects the fluid mass. Later we will see that vorticity is also advected along \(\bm{\mathcal{U}}\). 

The  explicit form of the flow field elementary follows from Eqs.~(\ref{4},\ref{6},\ref{8}) and its definition \eq{161} 
\begin{align}
 \la{17} 
    \bm{\mathcal{J}} 
    &= \rho\bm v- 2\mu^{\sA}(\bm\omega+\bm B)- {\bm E}_{\sA}\times m\bm v
 \\
    &= \rho\bm v-\mu^{\sA}(\bm\omega+2\bm B)-\bm\nabla\!\times\!(\mu^{\sA}m
\bm v)\,,
\end{align}
where we denote \(\bm E^{\sA} =\bm\nabla\mu^{\sA}\) as an analog of the axial ``electric'' field. In the important case, when \(\mu^{\sA}\) is constant across the fluid, the term \(\int\mu^{\sA}\rho_{\sA} d\bm x=\mu^{\sA}Q_{\sA}\) in \eq{4}
is the Casimir invariant that does not affect the equations of motion. In
this case
\begin{align}
\bm{\mathcal{J}}_{\mu^{\sA}=\rm const} &= \rho\bm v- 2\mu^{\sA}(\bm\omega+\bm
B)\,,
\end{align} 
so that \(\bm{\mathcal{J}}\) differs from \(\rho\bm v\) by a divergence-free term.

\section{Vector current and its anomaly}

Applying the definition \eq{19} to the Hamiltonian \eq{13} we obtain the
explicit expression of the vector current in terms of the Eulerian fields $\rho,\bm v$ \footnote{We assume that the definition \eq{162}  remains intact even if the system is in the magnetic field. In more  general situations, the electric current may contain additional divergence-free micro-currents (aka nonminimal terms) triggered by the magnetic field.}
\begin{align}
    \bm j=\rho\bm v+ \bm E^{\sA}\!\times\! m\bm v\,.
\la{162}
\end{align}
If \(\mu^{\sA}\) is uniform $\bm E^{\sA}=0$ and the vector current does not deviate from \(\rho\bm v\).  Otherwise, the vector current \eq{162} receives a component normal to velocity, a feature reminiscent of the Hall effect.

From \eq{162} and \eq{17} we obtain the relation between the flow current
and the vector current
\begin{align}
 \la{216}
    \bm{\mathcal{J}}=\bm j- 2\bm\nabla\!\times\!(\mu^{\sA} m\bm v)-2\mu^{\sA}\bm B\,.
\end{align}  
Then, the continuity equation \eq{20}, written in terms of the vector current
reads yields the chiral anomaly
\eq{51}
\begin{align}
    \dot \rho+\bm\nabla\cdot \bm j=2\bm\nabla \mu^{\sA}\cdot\bm B\,.
 \la{251}
\end{align}
Hence, from the perspective of the fluid dynamics, the chiral anomaly \eq{51} is yet another form of the continuity equation emphasizing the difference between the vector current \(\bm j\) and the advection flux \(\bm{\mathcal{J}}\) given by \eq{216}.

\section{Euler equation \la{E}} 

We now proceed with the Euler equation. We compute
\begin{align}
    \dot{\bm \pi}(\bm x)\!&=\!\{\bm \pi(\bm x), \mathscr{H}\}\! 
    =\int \left[\{\bm \pi(\bm x), \rho\}\frac{\delta \mathscr{H}}{\delta\rho} +\{\bm \pi(\bm x),  \bm\pi\}\frac{\delta \mathscr{H}}{\delta\bm\pi}\right]\,,
 \label{pdot}
\end{align} 
where the integration goes over the omitted argument. With the help of the Poisson brackets \eq{160} and the definitions (\ref{23},\ref{161}) we obtain the Euler equation
\begin{align}
    (\dot{\bm\pi}-\bm\nabla\pi_0)=\bm{\mathcal{U}}\times(\bm \nabla\times\bm\pi)\,.
 \la{29}
\end{align}      
If there is no chiral coupling   $\mu_{\sA}=0$,  the flow vector field \(\bm{\mathcal{U}}\) equals \(\bm v\) and \eq{29} becomes the familiar Euler equation for the perfect charged fluid \(m\dot{\bm v}-\bm v\!\times\!\bm\omega +\bm\nabla(\tfrac
m2\bm v^2+w)=\bm E+\bm v\!\times\! \bm B\).   

The form \eq{29} of the Euler equation is the basis of the geometric interpretation of the Euler hydrodynamics. It emphasizes the  Helmholtz law of advection of vorticity along the flow vector field \(\bm{\mathcal{U}}\). It is instructive to write the Euler equation in the space-time covariant form. Let us  introduce a space-time 4-momentum
\be
 \la{218}
    \pi_\alpha=(\pi_0,\bm\pi)
\ee 
and the canonical vorticity 2-form 
\be 
    \Omega_{\alpha\beta} =\p_\alpha\pi_\beta-\p_\beta\pi_\alpha\,.
\ee 
The vector field \(\bm{\mathcal{U}}\) relates the components of the vorticity so that \eq{29} can be written as \(\Omega_{0i}={\mathcal{U}}^j\Omega_{ji}\). This means that the flow advects the vorticity along the vector field \(\bm{\mathcal{U}}\) (the Helmholtz law). In the space-time notations \(\mathcal{J}=(\rho,\bm{\mathcal{J})}\), the Euler equation receives the form known in relativistic hydrodynamics as the Lichnerowitz equation \cite{rieutord2006introduction}
\begin{align}\la{Le}
        \mathcal{J}^\alpha\Omega_{\alpha\beta}=0\,.
\end{align}
In particular, this form of the Euler equation suggests that $\pi$ is a space-time covector, and the advection 4-current $\mathcal{J}$ is the space-time vector.

\section{ Momentum-energy equations  and the Lorentz force }\la{LF}

An insight about forces the reservoir exerts on the fluid comes from the equations for the fluid momentum $\rho{\bm v}$ and the energy density \(\mathcal{E}=\tfrac 12 m\rho  v^2+\varepsilon\) of an isolated fluid. We obtain them by transforming the Euler equation \eq{29} with the help of (\ref{20},\ref{17})  
\begin{align}
 \begin{cases}
    &\p_t(\rho{\bm v})+\p_i( v_i\rho\bm v) +\bm\nabla P= \rho\bm E+\bm j\times \bm B +\rho_{\sA} {\bm E}_{\sA}\,,
 \la{34} \\
    &\p_t\mathcal{E}+\bm\nabla\!\cdot\!\bm j_{\mathcal{E}}=\bm E\cdot\bm j+{\bm E}^{\sA}\cdot\bm {\bm j}_{\sA}\,,
 \end{cases}
\end{align} 
where $P=\rho w-\varepsilon$,\, is the pressure  and $\bm{j}_{\mathcal E}=(w+mv^2/2)\rho\bm{v}$ is the energy flux of the isolated fluid and 
\begin{align}
    \bm j_{\sA} =-\Phi(\bm\omega+2\bm B)\,-m\bm v\!\times\!\Big((\bm\omega+\bm B)\!\times\bm{\mathcal{U}}\Big)+\bm{E}\times m\bm{v}\,,
 \la{2500}
\end{align} 
In the next section, we show that \eq{2500} is the axial current of \eq{144}. The l.h.s. of these equations are the space-time divergence  of the momentum and the energy flux  of the isolated fluid, the r.h.s. are forces exerted by the reservoirs and electromagnetic field.

The highlight of these equations is that the vector current $\bm j$ defined by \eq{19} and given explicitly by \eq{162} enters
the Lorentz force in the right hand side of \eq{34}. Furthermore, Eq.~\eq{34} suggests that (i) \(\bm E^{\sA}\) acts on the fluid axial current \((\rho_{\sA},\bm{j}_{\sA})\) similarly to the action of the electric field \(\bm E\) on the vector current \((\rho,\bm j)\),  and that (ii) the vector current \(\bm j\) \eq{162} defining the Lorentz force is, indeed, the electric current defined by \eq{19}.

\section {Axial-current and its anomaly\la{sec25}} 
In a similar manner we derive the axial anomaly equation \eq{144}. With the help of the brackets (\ref{333},\ref{252}) we compute \(\p_t \rho_{\sA}=\{\rho_{\sA},\,\mathscr{H}\}\).  Proceeding similarly to \eq{pdot} and using the notations (\ref{23},\ref{161}) we obtain
\begin{align}
    \p_t \rho_{\sA}(\bm x) =\int\Big(\left\{\rho_{\sA} (\bm x),  \bm\pi\right\}\cdot\bm{\mathcal{J}} -\left\{\rho_{\sA} (\bm x),  \rho\right\}\pi_0\Big)\,,
 \label{rhoAdot}
\end{align}   
[the integration goes over the omitted argument]. Then with the help of \eq{333} and \eq{252}, we obtain the axial anomaly equation \eq{144} and the hydrodynamic expression for the axial current. The latter is given by \eq{2500} and coincides with the current ${\bm j}_{\sA}$ entering the evolution equation \eq{34} for the energy density.

Finally, with the help of the Euler equation \eq{29} and using the notation
\begin{align}
 \la{223}
    p_\alpha:=\pi_\alpha-A_\alpha=\left(-w[\rho]-\tfrac 12m\bm { v}^2,m\bm v\right)\,,
\end{align}
we write the chirality 4-current \(j_{\sA}=(\rho_{\sA},\bm j_{\sA})\)
in a space-time covariant form previously obtained in \cite{abanov2022axial}
\begin{align}\la{224}
    j_{\sA}^\alpha=\e^{\alpha\beta\gamma\delta}p_\beta(\p_\gamma p_\delta+F_{\gamma\delta})\,,
\end{align}
where \(F_{\alpha\beta}=\p_\alpha A_\beta -\p_\beta A_\alpha\) is the electromagnetic field tensor.

\section{Beltrami flow} 

Now we are equipped to discuss stationary flows. We assume the external fields  are either time-independent, or depend on time adiabatically. The stationary (or steady) flows give extrema to \(\mathscr{H}\). The variation of \(\mathscr{H}\) with respect to the fluid density is the Bernoulli function \eq{23}. The variation with respect to the momentum is the flow vector field \eq{161}. In the stationary flow, whose density and velocity we denote by \(\bar \rho,\bm{\bar v} \), both these variations should be zero which gives the conditions
\begin{align}
    \begin{cases}
    \pi_0=0:\quad & w[\bar\rho]+\tfrac 12m \bm { \bar v}^2=\mu,
 \\
    \bm{\mathcal{J}}=0:\quad&\rho\bm {\bar v} =2\mu^{\sA}(\bm{\bar \omega}+\bm
B) +\bm E^{\sA}\times m \bm{\bar v}\,.
 \la{350}
    \end{cases}
\end{align}
The first line is the Bernoulli condition. The second defines the generalized Beltrami flow in the magnetic field. Together they determine \(\rho,\rho_{\sA}\) as functions of \(\mu,\mu^{\sA}\). For example, at \(\bm B=0\) we have the relation \(\mu^{\sA} {\bar \rho}_{\sA}=(\mu-w[\bar \rho])\bar\rho\).

We recall that in the absence of magnetic field, the velocity of the generalized Beltrami flow is an eigenfunction of the \(\rm curl\) operator 
\be
    {\rm curl}\,\bm {\bar v}=\kappa\bm {\bar v}\,.
\ee 
Extension by the magnetic field amounts to the shift by the Larmor frequency \({\rm curl}\,\bm v\to{\rm curl}\,\bm v+\bm B/m\). In our case \({\rm\ curl}\, \bm v=\tfrac 1{\sqrt{\mu^{\sA}}}\bm\nabla(\sqrt{\mu^{\sA}}\times\bm v)\) and \(\kappa=\rho/2m\mu^{\sA}\). Locally Beltrami flow describes a helix along a vortex line. The pitch of the helix of the order \(1/\kappa\) (see, e.g., \cite{dritschel1991generalized}).

A particularly interesting case is the Beltrami flow proper, when  \(\kappa=\text{const}\). In this case, the Beltrami flow is incompressible as \({\rm div}\,\bar{\bm v}=0\), where \({\rm div}\,\bm v=\tfrac 1{\sqrt{\mu^{\sA}}}\bm\nabla\cdot(\sqrt{\mu^{\sA}}\,\bm v)\). An important feature of the streamlines of the incompressible  Beltrami flow is that they are not integral, and some are known to be chaotic \cite{arnold2008topological}. A similar situation  occurs in some compressible (generalized) Beltrami flows \cite{morgulis1995compressible} (see also \cite{enciso2016beltrami} for conditions when compressible solutions exist). We do not elaborate on this aspect further but remark that a possibility of chaotic advection of particles (Lagrangian turbulence) in the context of fermionic QFT seems to be an unexplored aspect of anomalies.

\subsection*{Transport current of  the Beltrami flow\la{Sec29}} 

Let us evaluate the vector and the axial current on the Beltrami flow. This way we obtain the equilibrium value of the currents. Combining
(\ref{162},\ref{2500},\ref{350}) we obtain  
\begin{align}
 \label{233}
    &\bar{\bm j}=2\mu^{\sA}\bm B+2\bm\nabla\!\times\! (\mu^{\sA} m\bm{\bar v}) \,,
    \quad 
    \bar{\bm{j}}_{\sA}=2\mu\bm B+\bm\nabla\!\times \!(\mu m\bm{\bar v})\,.
\end{align}
We notice that both currents \eq{233} consist of the transport part 
\be
    \bm{j}_{\rm trans}=2\mu^{\sA}\bm B\,,
    \quad 
    \bm{j}_{\sA\, \rm trans}=2\mu\bm B\,,
 \la{233tr}
\ee 
and a microscopic current, the curl
\be
    \bm j_{\rm micro} =2\bm\nabla\!\times\! (\mu^{\sA} m\bm{\bar v}) \,,
    \quad 
    \bm j_{\sA\, \rm micro} =\bm\nabla\!\times \!(\mu m\bm{\bar v})\,.
 \la{233mag}
\ee 
If the chemical potentials are constant, the vector and axial currents currents are\footnote{\label{note7} Formulas similar to \eq{38} appear in various hydrodynamic models  (other than Euler hydrodynamics) aimed to be consistent with the chiral anomaly. There the term proportional to the magnetic field and the term proportional to the vorticity in the vector current   is dubbed by the  `chiral magnetic effect' and the 'chiral vortical effect', respectively (see, e.g., 
\cite{kharzeev2011testing} and references therein). Similar terms in the axial  current  are sometimes
referred to as the `chiral separation effect' and the 'chiral vortical separation effect', respectively \cite{chernodub2021thermal}. The coefficients vary
across the literature.}
\begin{align}
    \bar{\bm j}=&2\mu^{\sA}(\bm B+\bm\omega)\,,
    \quad {\bar{\bm{j}}_{\sA}}=2\mu\left(\bm B+\tfrac 12\bm\omega\right)\,,
    \quad
    \mu,\mu^{\sA}=\rm const\,.
 \la{38}\
\end{align}

The micro-currents describe the stationary movement of electric and axial charges localized on vortex loops. In contrast, the transport currents transfer hydrodynamics substances along magnetic lines propagating across the system. 

As we remarked in the Sec.~\ref{sec25} one can trace the origin of the transport current to the central extension of the Poisson algebra \eq{333}. The expressions (\ref{233}-\ref{38}) are the main results of this work. We have shown that an equilibrium state of the fluid coupled to the chiral reservoir (i) is a Beltrami flow and that (ii) the Beltrami flow transports both electric and the chirality charges in a way identical to the anomaly-driven flows of Dirac fermions.

\section{Flows in an external axial potential \label{flows}} 

We conclude the paper by extending the coupling to the chiral reservoir from $\mu^{\sA}$ to a full axial 4-potential  \( \mu^{\sA}\to A_\alpha^{\sA} =(\mu^{\sA},\bm A^{\sA})\). The extended coupling brings various currents to space-time-covariant forms and further clarifies the parallels between the fluid dynamics and Dirac fermions in the axial field. An economical way to obtain the results is to use almost obvious covariant extensions of the formulas of the main text. We present the main expressions here and publish a detailed derivation separately.

The covariant extension is based on the main property of the Euler hydrodynamics: the fluid  momentum
\(p_\alpha=\pi_\alpha-A_\alpha\), and the advection current \(\mathcal{J}^\alpha\), are \(\mathbb{R}^4\) covector and vector, respectively, regardless of the space-time structure. For that reason the formulas below are equally valid in the relativistic case.  

We start from the currents. The axial current \eq{224} does not depend on the axial potential and has already appeared in the space-time covariant form. In turn, the vector current does not depend on the vector potential. The advection 4-current is the covariant extension of (\ref{216}). Together the currents read 
\begin{align}
 \begin{cases}
    j^\alpha=\rho u^\alpha+^\star \!\! F_{\sA}^{\alpha\beta}p_\beta,\quad
&
    \\
    j_{\sA}^\alpha =\epsilon^{\alpha\beta\gamma\delta}p_\beta(\p_\gamma
p_\delta\,+F_{\gamma\delta}),\quad &
\\
    \mathcal{J}^\alpha=j^\alpha + 2\,{}^\star \! F^{\alpha\beta}A^{\sA}_\beta
+2\epsilon^{\alpha\beta\gamma\delta} \p_\gamma (A^{\sA}_\beta  p_\delta).
 \end{cases}\la{41}
\end{align}
In these formulas we denote the 4-velocity
\(u^\alpha=(1,\bm v)\). Also, \(^\star\!
F^{\alpha\beta}=\tfrac 12\epsilon^{\alpha\beta\gamma\delta} F_{\gamma\delta}\) is the dual field tensor,  \(F_{\alpha\beta}=\p_\alpha
A_\beta-\p_\beta A_\alpha\) and we use similar notations for the axial field tensors.
 
As we mentioned in the Sec.~\ref{E}, the covariant form of the Euler equation appears as the Lichnerowitz equation \eq{Le}. We reproduce it here
\begin{align}\begin{cases}\nn
    \mathcal{J}^\alpha(\p_\alpha\pi_\beta-\p_\beta\pi_\alpha)=0,\\
    \p_\alpha \mathcal{J}^\alpha=0\,.
    \end{cases}
\end{align}
The covariant form of the divergence of the stress tensor extends the equation \eq{34} and the divergence of the vector current extends  the formula \eq{251}. Together they   are given  by
\begin{align}
    \begin{cases}
    \p_\beta T_\alpha^\beta =F_{\alpha\beta}\,j^\beta+ F^{\sA}_{\alpha\beta}\,j^\beta_{\sA}\,,
 \\
    \p\!\cdot\! j=F\!\cdot\!{}^\star\! F^{\sA}=2(\bm E^{\sA}\!\cdot \!\bm B+\bm B^{\sA}\!\cdot \!\bm E)\,,
 \\
    \p\!\cdot\!j_{\sA}=\tfrac 12F\!\cdot\!^\star\!F=2 \bm E\!\cdot \!\bm B \,.
    \end{cases}
 \la{40}
\end{align}
The equations \eq{41} and \eq{40} retain their form for the relativistic fluid. For the non-relativistic fluid the components of the momentum-stress-energy tensor in \eq{40} are:  the energy density \(-T_0^0=\tfrac 12m\rho\bm v^2+\varepsilon[\rho]\), the energy flux \(-T^i_0=\rho v^i (\tfrac 12 mv^2+w)\), the momentum density \(T^0_i=m\rho v_i\), and the momentum flux  \(T_i^j=m\rho v_iv^j+P\delta_i^j\). For a relativistic fluid $T_{\alpha}^\beta=\rho\, p_\alpha u^\beta  +P\delta_{\alpha}^\beta$, and the entries in \eq{41} are the relativistic 4-velocity  $u^\alpha=(1,\bm v/c)/\sqrt{1-\bm v^2/c^2}$  and the  4-momentum $p_\alpha = (mc+c^{-1}w)u_\alpha$. 

We comment that starting from the  Eqs.\eq{40}  and given momentum-stress-energy tensor, one could derive the expressions for the currents \eq{41} as a consistency of \eq{40}.  This argument had been explored in \cite{son2009hydrodynamics} for the relativistic fluid.\footnote{\label{note5}  We comment on the formal relation between Eqs.~(\ref{41},\ref{40}) and the formulas of Refs.~\cite{son2009hydrodynamics,monteiro2015hydrodynamics}.
Equations  $\p_\beta T_\alpha^\beta =F_{\alpha\beta}\,j_{[1]}^\beta$
and  $\p_\mu j_{[1]}^\mu=C\bm E\cdot \bm B$ of  \cite{son2009hydrodynamics,monteiro2015hydrodynamics}
follow from \eq{40} in a particular background when  the vector and the axial  potentials are set proportional  $A^{\sA}=CA/6$.  Under this condition $j_{[1]}$ of Ref.~\cite{son2009hydrodynamics} is  a particular  linear combination  $j+Cj_{\sA}/6$ of the currents given by \eq{41}.}

\section{Summary}

Summing up, we have shown that anomaly-based properties of QFT with Dirac fermions coupled to the axial vector potential are identical to the kinematic properties of the Euler fluid coupled to the reservoir, which maintains the fluid helicity. The result is well captured by Eqs.~(\ref{41},\ref{40}). From the perspective of fluid dynamics, the equation for the stress tensor (the first equation in \eq{40}) is the complete equation of motion. Once the stress tensor and the currents are expressed through the fluid density and the velocity as in \eq{41}, the divergence of the currents (the last two equations in \eq{40}) follow. Now let us look on \eq{40} from the perspective of Dirac fermions. The first equation in \eq{40} is valid for the classical  Dirac equation, or quantum alike. However, the divergences of the currents (the last two equations) are of a quantum nature. They require a short-distance regularization which preserves the gauge invariance. That regularization yields the anomaly in the divergence of the currents. Once the divergence is determined, the equations must be completed by constitutive relations, for example, expressing the stress tensor and the axial current in terms of the vector current. We have shown that the Euler fluid provides such constitutive relation in the form of Eqs.~\eq{41}. 
Another result of our work is a close parallel between the Beltrami flow in fluid dynamics and the ground state of the Dirac fermions. The Beltrami flow stands out as a fascinating property of fluid dynamics. In turn, the quantum anomalies might be understood from the properties of the ground state of the Dirac fermions (perturbed by the external fields). A close parallel between these objects of seemingly unrelated fields seems noteworthy. Interpretation and consequences of the chaotic nature of the Beltrami flow in the quantum field theory are especially interesting. 

Equations of motions \eq{40} and the hydrodynamic expressions for the vector and the axial currents (\ref{41}) are gauge invariant with respect to both the vector \(A\to A+\p\varphi\) and the axial \(A^{\sA}\to A^{\sA}+\p\varphi^{\sA}\) gauge transformations. However, the advection current   $\mathcal{J}$ given by the third line of \eq{41} and the Beltrami flow defined by the condition \(\bm{\mathcal{J}}=0\) are not invariant under the axial gauge transformation. As a result, the quantities evaluated on the stationary flow \eq{350} do not generally possess the axial gauge symmetry. Despite the axial gauge symmetry of equations of motion, the symmetry does not hold at the equilibrium state (or the ground state in QFT). This is the essence of the anomaly: the (axial) symmetry of equations of motions is not a symmetry of their solutions. This property gives rise to the main effect we discussed in the paper. The equilibrium state (the Beltrami flow) possesses a non-vanishing mass flux described in Sec.~\ref{29}. Even if the axial vector potential is a pure gauge, so that the axial field is a gradient  \(A_\beta^{\sA}=(2\pi)^{-1}\p_\beta\Theta\), hence drops out of the equations of motions \eq{40}, the advection current \(\mathcal{J}\), hence the state of  equilibrium \(\bm{\mathcal{J}}=0\) depends on \(\Theta\). As a result, the equilibrium state transports the mass/charge with an adiabatic change of the theta-angle as $\bm j_{\rm trans}=(1/\pi)\bm B\dot \Theta$. The amount of charge crossing the surface element $d\sigma$ when $\Theta$ changes by $2\pi$ is $2Bd\sigma$, i.e., twice the magnetic flux threading the surface.

\acknowledgments
We thank A.~Alekseev, A.~Cappelli, G.~Monteiro and V.~P.~Nair for their interests in this study and useful discussions. We thank G.~Volovik and B.~Khesin for comments and bringing the Refs.~\cite{bevan1997momentum,salomaa1987quantized,khesin1989invariants} to our attention.  The work of P.B.W. was supported by the NSF under the Grant NSF DMR-1949963. The work of A.G.A. was supported by NSF DMR-2116767.


\bibliographystyle{JHEP.bst}
\bibliography{helicity.bib}

\end{document}